\documentclass[preprint,floats,aps,epsfig,nofootinbib,amssymb]{revtex4-1}
\usepackage{mathrsfs}

\usepackage{slashed}
\usepackage{graphicx,color}
\usepackage{subfigure}
\usepackage{dcolumn}
\usepackage{bm}
\usepackage{subfig}
\usepackage{graphicx}
\usepackage{amssymb}
\usepackage{stackrel}
\usepackage{pgfplots}
\usetikzlibrary{positioning}
\usetikzlibrary{snakes}
\begin{document}

\title{Neutrino Portal via Loops}
%\title{Loop Portal, screening mass and neutrino-antineutrino oscillation}

\author{Wei Chao}
\email{chaowei@bnu.edu.cn}

\affiliation{Center for Advanced Quantum Studies, Department of Physics, Beijing Normal University, Beijing, 100875, China}
\vspace{3cm}

\begin{abstract}

One loop correction to the dark matter (DM) annihilation cross section is usually assumed to be sub-dominate in calculating the DM relic density. In this paper, we propose a scenario where DM  freezes out mainly via loops in the neutrino portal. For a scalar DM annihilating into neutrino via a triangle diagram, or a Majorana DM annihilating into neutrino via a box diagram, the observed DM relic density can be generated if  the mediator is Majorana fermion. We further work out the neutrino Debye mass and neutrino oscillation probabilities in the dark halo.  Numerical results show that impactions of DM to these physical observables are negligible on the Earth,  however the analytical results can be applied to evaluate neutrino properties in a dense DM environment.

\end{abstract}

\maketitle
\section{Introduction}

Astrophysical observations have confirmed the existence of dark matter (DM)~\cite{Aghanim:2018eyx}, which cannot be addressed by the minimal Standard Model (SM) of particle physics. Of all the SM particles, neutrinos are most similar to DM in nature. In fact, neutrinos themselves are candidates for hot DM, which however cannot explain the formation of individual galaxies after the big bang. To explain the small-scale structure in the early Universe, we must rely on cold DM.  What is the mass of the cold DM? how it couples to the SM particles? and what is the thermal history of the cold DM in the early Universe?  are three unsolved mysteries. For the past decades, many DM models have been created with mass ranging from $10^{-20}$~{\rm eV} to $10^{55}~{\rm GeV}$, of which, weakly interacting massive particle~\cite{Goldberg:1983nd,Ellis:1983ew,Jungman:1995df,Servant:2002aq,Cheng:2002ej,Bertone:2004pz}, sub-GeV DM~\cite{Essig:2011nj}, axion~\cite{Marsh:2015xka,Preskill:1982cy,Abbott:1982af,Dine:1981rt} and primordial black hole~\cite{Carr:2016drx} are typical DM candidates with abundant physical signals.

In this paper, we focus on the thermal history of the cold DM in the early Universe.
It is well-known that DM can be produced thermally via the freeze-out or freeze-in mechanism~\cite{Hall:2009bx}, or non-thermally via the misalignment mechanism~\cite{Preskill:1982cy,Abbott:1982af,Dine:1981rt}, depending on the dynamics of the hidden sector. Non-thermal production mechanism is usually related to the phase transition, during which epoch the energy stored in the scalar potential false vacuum can be transferred to the DM. In the freeze-in scenario, DM is produced by the annihilation or decay of heavier particle, but the DM interaction rate is too small to thermalize.  The freeze-out scenario requires a large enough interaction rate. If DM freezes out relativistically, then its relic density is simply $\Omega h^2 =(78/g^*_s)(m_{\rm DM}/{\rm KeV}) $~\cite{Hambye:2020lvy,Frigerio:2011in}, with $g^s_*$ the number of degrees of freedom contributing to entropy at DM decoupling.  Alternatively, if DM freeze-out non-relativistically, the relic density depends only on the annihilation cross section, which takes the value $\langle \sigma v \rangle \sim 10^{-26}{ \rm cm^3\cdot s^{-1}}$~\cite{Bertone:2004pz}.
 
We propose an interesting  possibility where DM is thermalized via tree-level interactions but non-relativistically freeze-out via high order processes. Usually these loop effects are negligibly small and cannot solely fit to a correct relic abundance when the tree-level annihilation process is kinematically forbidden. We find that the one-loop contribution can be enhanced in the neutrino portal whenever the mediator is a Majorana fermion.  To illustrate this point, we propose two sub-GeV DM models: a scalar DM model, which freezes out via a triangle diagram, and a Majorana DM which freezes out via a box diagram. In both scenarios, the loop integrals are finite and  large enough give rise to a correct relic abundance. Considering that thermal correction to the relic density is only proportional to ${\cal O}(T^4)$~\cite{Beneke:2014gla,Beneke:2016ghp}, the result of loop calculation at the  zero-temperature is credible. 

It has been shown that active neutrino may acquire a sizable nonzero mass from a long range potential sourced by DM distributions~\cite{Davoudiasl:2018hjw,DAmico:2018hgc}. We evaluate the neutrino Majorana mass induced by the scalar DM. It shows that the induced neutrino mass in the loop-portal is proportional to the DM density, and is negligibly small on the Earth as the local DM density is too low. We further investigate the matter effect of neutrino oscillations induced by the DM. For the first time, we derive the  formulae of DM effect in neutrino-antineutrino oscillation probability, which is applicable to the investigation of neutrino oscillations with large Debye mass.  For loop-portal scenarios, it is unlikely to test the DM effect in neutrino oscillation experiments in the solar system.  Our results, in addition to the conclusions of Refs.~\cite{Choi:2019zxy,Liao:2018byh,Capozzi:2018bps} (and these from references cited therein), clarify the physical picture of neutrino oscillations in the dark halo.

The remaining of the paper is organized as follows: In section II we introduce the loop-portal in detail and calculate the DM relic density. Section III is devoted to the study of neutrino mass induced by the DM. In section IV, we study the DM effect in neutrino oscillations. The last part is concluding remarks.

\section{Loop-portal}

In this section we present two models of neutrino-portal via  loops: the scenario A for Majorana DM $\chi$ and the scenario B for  scalar DM $\varphi$. They are both two-component DM models, of which the heavier DM component will dominantly annihilate into the lighter one resulting  in negligible relic density. We assume the heavier DM component is a complex scalar $\Phi$, which couples to the right-handed Majorana neutrino $N_R^{}$ and the left-handed active neutrinos $\nu_L$: $ {\cal L}_{int} = Y_\nu^{} \Phi \overline{ \nu_L^{} }  N_R^{}  + {\rm h.c.} $, with $Y_\nu$ the Yukawa coupling.\footnote{Such an interaction is natural if $N_R^{}$ is the neutral component of a vector-like lepton doublet which gets non-zero Majorana mass via a seesaw-like mechanism.  Alternatively, if $ \Phi$ mixes with the neutral component of an inert Higgs doublet $\eta$ via the interaction $ V\sim \Delta \Phi H^\dagger \eta$, and $\eta$ couples to $N_R$  and  left-handed lepton doublet $\ell_L$ ($ Y_\eta \overline{\ell_L^{}} \tilde\eta N_R^{} $), one may also get this Yukawa interaction.}
Then Lagrangians for scenarios A and B can be written as
\begin{eqnarray}
-{\cal L}_{\rm int}^A &=& {1\over 2 } m_\chi \overline{\chi_L^{} } \chi_L^C +  Y_\chi^{} \overline{\chi_L^{} } \Phi N_R^{} +Y_\nu^{} \overline {\nu_L^{}}   \Phi N_R^{}  + {\rm h.c.}  \\
-{\cal L}_{\rm int}^B &=&{1\over 2 } m_\varphi^2 \varphi^2 + {1\over 2}\lambda \Phi^2 \varphi^2+ Y_\nu^{} \overline {\nu_L^{}}   \Phi N_R^{}  + {\rm h.c.} 
\end{eqnarray}
where $m_\chi$ and $m_\varphi$ are masses of $\chi$ and $\varphi$ respectively,  $Y_\chi$ is the Yukawa coupling and $\lambda$ is the quartic coupling.  Interactions between the SM Higgs and new scalar singlets, as well as quartic coupling between $\varphi^2$ and  $\Phi^\dagger \Phi$ bilinears, are neglected for simplicity. Throughout this paper we  focus on the thermal dynamics of sub-GeV DMs and neglect  the active neutrino mass. 

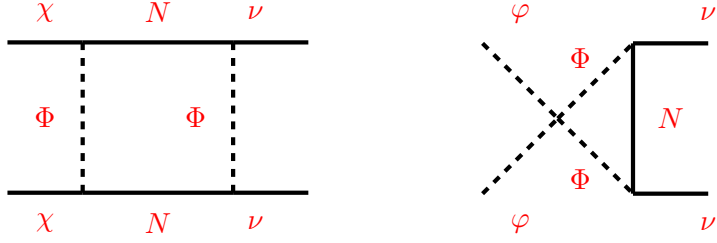
\begin{figure}[t]
\begin{center}
\begin{tikzpicture}
\draw[-,ultra thick] (-1,0)--(0,0);
\draw[-,ultra thick] (-1,2)--(0,2);
\draw[-,ultra thick,dashed] (0,0)--(0,2);
\draw[-,ultra thick] (0,2)--(2,2);
\draw[-,ultra thick] (0,0)--(2,0);
\draw[-,ultra thick] (2,2)--(3,2);
\draw[-,ultra thick] (2,0)--(3,0);
\draw[-,ultra thick,dashed] (2,0)--(2,2);
\node[red, thick] at (2.3,2.4) {$\nu$};
\node[red, thick] at (2.3,-0.4) {$\nu$};
\node[red, thick] at (1.5,1) {$\Phi$};
\node[red, thick] at (-0.5,1) {$\Phi$};
\node[red, thick] at (1,2.4) {$N$};
\node[red, thick] at (1,-0.4) {$N$};
\node[red, thick] at (-0.5,2.4) {$\chi$};
\node[red, thick] at (-0.5,-0.4) {$\chi$};
\end{tikzpicture}
\hspace{2cm}
\begin{tikzpicture}
\draw[-,ultra thick,dashed] (-1,0)--(0,1);
\draw[-,ultra thick,dashed] (-1,2)--(0,1);
\draw[-,ultra thick,dashed] (0,1)--(1,2);
\draw[-,ultra thick,dashed] (0,1)--(1,0);
\draw[-,ultra thick] (1,2)--(2,2);
\draw[-,ultra thick] (1,0)--(2,0);
\draw[-,ultra thick] (1,0)--(1,2);
\node[red, thick] at (2,2.4) {$\nu$};
\node[red, thick] at (2,-0.4) {$\nu$};
\node[red, thick] at (1.5,1) {$N$};
\node[red, thick] at (0.3,1.8) {$\Phi$};
\node[red, thick] at (0.3,0.2) {$\Phi$};
\node[red, thick] at (-0.5,2.4) {$\varphi$};
\node[red, thick] at (-0.5,-0.4) {$\varphi$};
\end{tikzpicture}

\caption{Feynman diagrams for the annihilation of Majorana dark matter(left-panel) and scalar dark matter (right-panel). }\label{feynmand}
\end{center}
\end{figure}

In scenario A, both $\chi$ and $\Phi$ are DM candidates stabilized by a $Z_2$ symmetry. 
Assuming the mass hierarchy $m_N> m_\Phi > m_\chi$, $\Phi$ annihilates into $\chi$ in the early Universe, resulting in negligible relic abundance for a sizable Yukawa coupling $Y_\chi$, while $\chi$ can only annihilate into neutrinos via the Box diagram given in the left-panel of the  Fig.~\ref{feynmand}.  The thermal average of the reduced annihilation cross section can then be written as
\begin{eqnarray}
\sigma v_{\chi \bar \chi \to \nu \bar \nu} \approx  { m_\chi^2 {\cal I }^2 \over 32\pi m_\Phi^4 }
\end{eqnarray}
where ${\cal I}$ is the four point loop function proportional to the Passarino-Veltman scalar integral $D_0$~\cite{tHooft:1978jhc,Passarino:1978jh,Denner:1991kt}. 
To the leading order(neglecting the momentum of DM and neutrino in the propagators of heavy neutrino),  ${\cal I}$ takes the form:
\begin{eqnarray}
{\cal I} =  { Y_\nu^2 Y_\chi^2  \xi  \over 16\pi^2} \left\{{ (3+\xi+x ) \over (\xi-1)^2 \lambda^{1/2} (-x,\xi,1) } \log\left[ \lambda^{1/2} (-x,\xi,1) +x+\xi + 1 \over 2\sqrt{\xi}\right] \right.&& \nonumber \\ \left.+ {2 b_x\over \sqrt{x} (\xi-1)^2} \log\left[{2 b_x \sqrt{x} +x+2 \over 2}\right] + {\log(\xi) \over (\xi-1)^2}\right\}&&
\end{eqnarray}
where $\xi=m_N^2 /m_\Phi^2 $, $x=m_\chi^2 / m_\Phi^2$ and $b_x=\sqrt{1+x/4}$. The loop integral is calculated and simplified with the help of the Package-X~\cite{Patel:2015tea,Patel:2016fam}. There are neutrino-number-violating box diagrams, but their contribution to the annihilation are automatically cancelled.   

\begin{figure}
\centering
\includegraphics[width=0.42\textwidth]{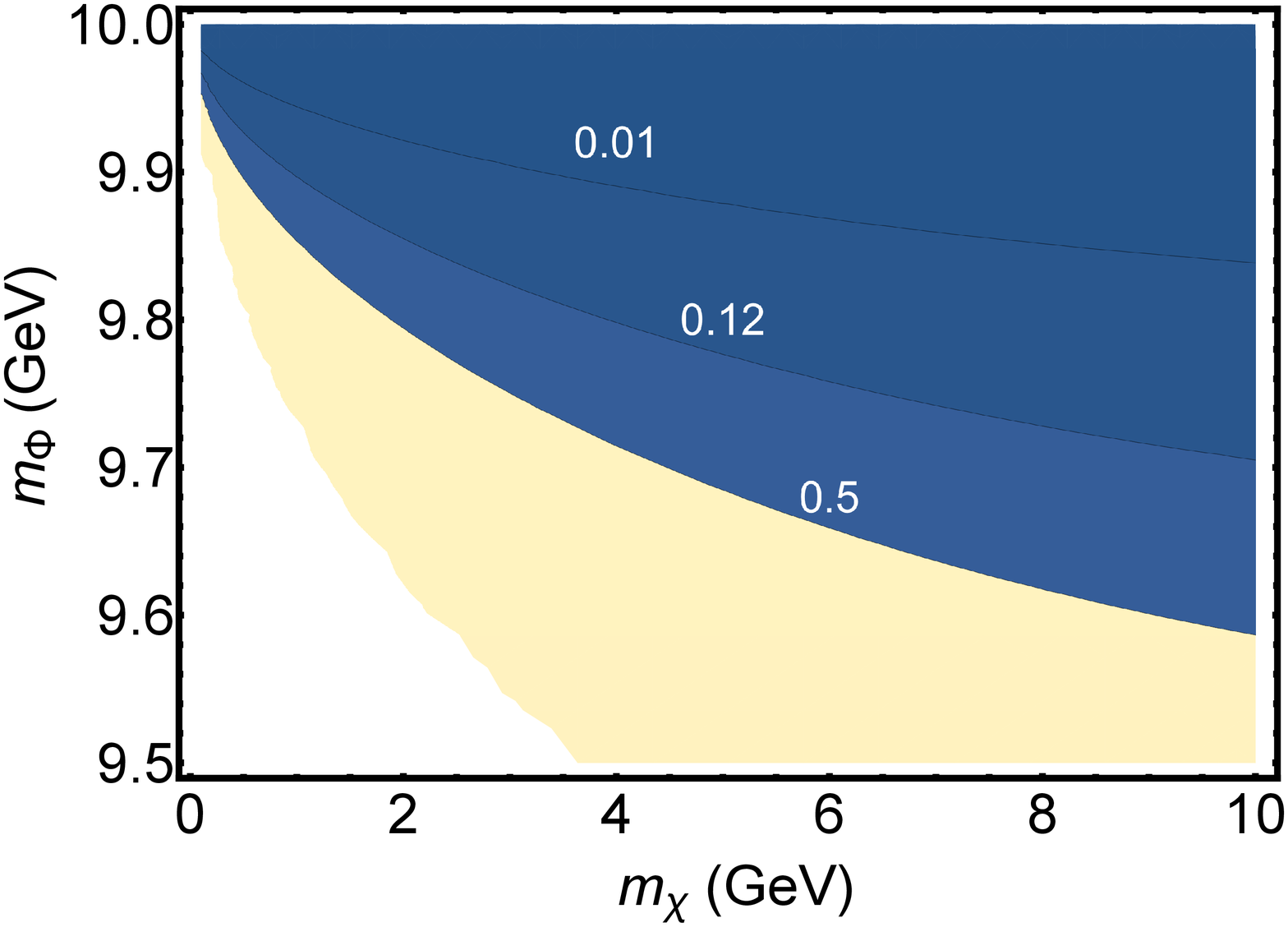}
\hspace{0.5cm}
\includegraphics[width=0.42\textwidth]{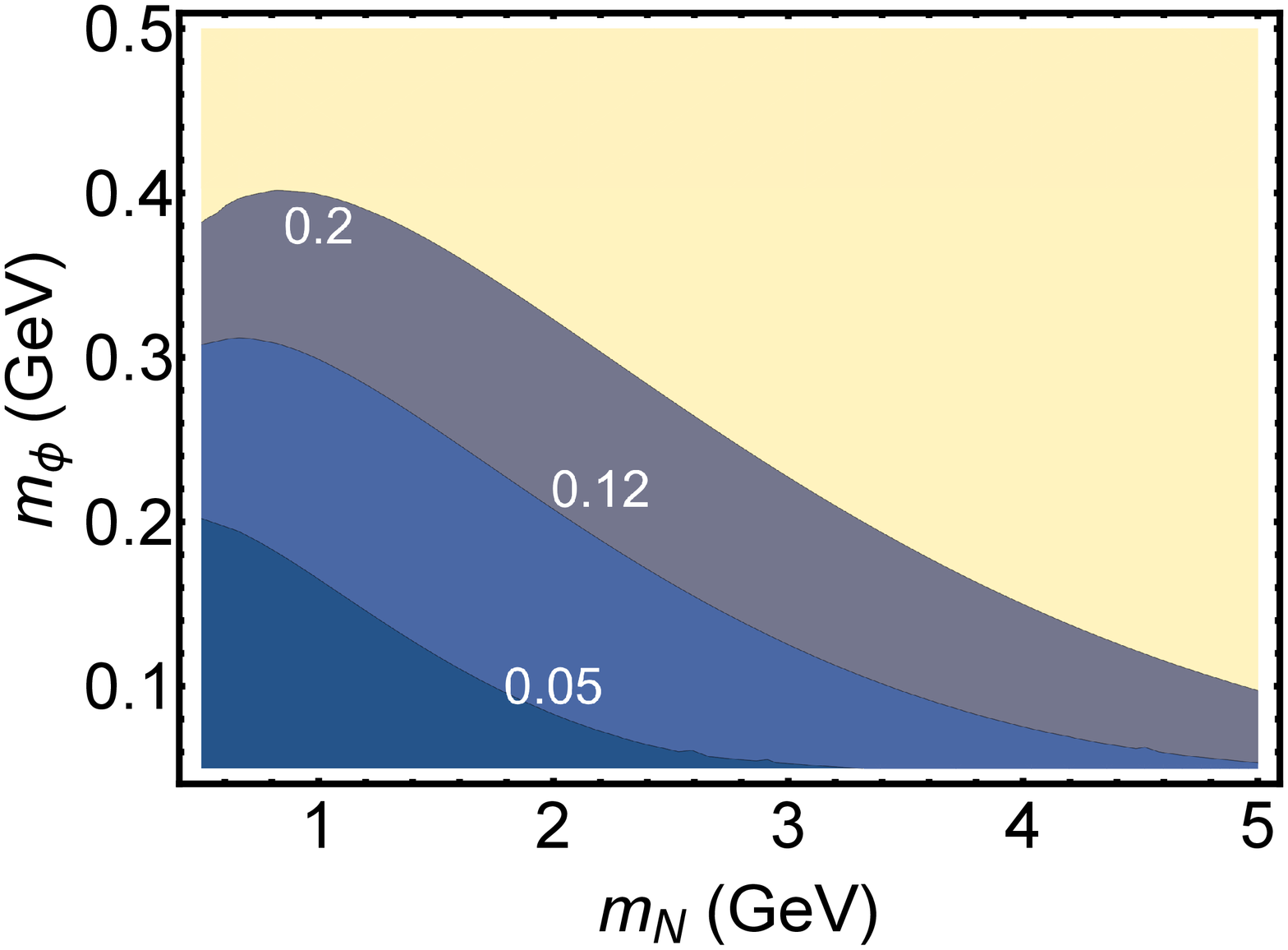}
\caption{ Left-panel: Contours of the relic density of $\chi$ in the $m_\chi-m_\Phi$ plane by setting $m_N=10~{\rm GeV}$, $Y_\chi=1$ and $Y_\nu=0.2$; Right-panel:The relic abundance of $\varphi$ in the $m_\phi-m_N$ plane, by setting $\lambda=Y_\nu=0.5$. Contours form the top to the bottom correspond to $\Omega h^2=0.2,~0.12$ and $0.05$, respectively.
} \label{fig:relicd}
\end{figure}

As an illustration, we show in the left-panel of the Fig.~\ref{fig:relicd} contours of the relic abundance of $\chi$ in the $m_\chi-m_\Phi$ plane by setting $Y_\chi=1$, $Y_N=0.2$ and $m_N=10~{\rm GeV}$.  Contours from the top to the bottom correspond to $\Omega h^2 =0.01, 0.12$ and $0.5$, respectively.  The lore from Tremaine and Gunn~\cite{Tremaine:1979we}, which is based on the Pauli exclusion principle,  sets a lower bound $m_{DM} >100~{\rm eV}$ on the mass of fermionic dark matter.  Besides, a lower bound for a cold DM mass is about $1~{\rm keV}$~\cite{Steigman:1984ac}.  We take $m_\chi >10~{\rm MeV}$ in the numerical simulation. Yukawa interactions in this scenario may contribute to active neutrino mass as well as neutrino-DM mixing through the so-called scotogenic mechanism~\cite{Farzan:2012sa} whenever there is a mass splitting between the CP-even and the CP-odd components of $\Phi$.  Alternatively, this constraint can be avoided for a degenerate $\Phi$.

In scenario B, both $\Phi$ and $\varphi$ are DM candidates, while the relic density of $\Phi$ can be small for a sizable quartic coupling $\lambda$.  $\varphi$  freezes out non-relativistically via a triangle diagram, as shown in the right-panel of the Fig. \ref{feynmand}, in which both neutrinos go out and the neutrino number is thus violated. The thermal average of the reduced annihilation cross section takes the form
\begin{eqnarray}
\langle \sigma v \rangle_{\varphi \varphi \to \nu \nu}\approx {1\over 8\pi M_N^2} {\cal F}^2 \left({M_\Phi^2  \over M_N^2}, {4M_\varphi^2\over M_N^2 }\right) \label{thermalaverage}
\end{eqnarray}
where 
\begin{eqnarray}
{\cal F} (\alpha, \beta) ={\lambda_1^{} \zeta^2 \over 16 \pi^2  } \int_0^1 dx {1\over 1- \alpha +\beta x} \left\{ \log [1-(1-\alpha) x] + \log[\alpha-\beta x(1-x)] \right \}.
\end{eqnarray}

The relic abundance of $\varphi$ is illustrated in the right-panel of Fig.~\ref{fig:relicd}, where we have set $\lambda =Y_\nu=0.5$. Contours from the top to the bottom correspond to $\Omega h^2 =0.2,~0.12$ and $0.05$, respectively. The relic density is insensitive to the dark matter mass as can be seen from the Eq.~(\ref{thermalaverage}). The relic density of $\Phi$ is about $\Omega h^2(\Phi)\sim10^{ -6}$ in this case.  An issue one may worry about is the thermal corrections in evaluating the loop diagram. It has been shown in Refs.~\cite{Beneke:2014gla,Beneke:2016ghp} that the leading finite temperature contribution is actually of the order ${\cal O} (T^4)$,  the soft and collinear temperature dependent divergences are cancelled, thus the zero-temperature calculation is applicable to this case.

%%%%%%%%%%%%%%%%%%%%%%%%%%%%%%%%%%%%%%%%%%%%%%%%%%%%%%%%%%%%%%%%%%%%%%%%%%%%%%%%%%%%%%%%%%%%%%%
%%%%%%%%%%%%%%%%%%%%%%%%%%%%%%%%%%%%%%%%%%%%%%%%%%%%%%%%%%%%%%%%%%%%%%%%%%%%%%%%%%%%%%%%%%%%%%%

In the following let us comment on possible signatures of loop-portal DM. Since DMs only interact with the heavier state $\Phi$, which is almost depleted at later times, direct detection is kinematically forbidden even if the heavy Majorana neutrino possesses other interactions. Alternatively, if $\Phi$ possesses the Higgs portal interaction, the DM-nucleon scattering may arise at the one-loop level, which has been shown to be small as compared with current constraints put by DM direct detection experiments~\cite{Chao:2019lhb,Li:2020pfy}. High luminosity fixed target neutrino experiments have been identified as good facilities of detecting sub-GeV DM~\cite{deNiverville:2012ij,Morrissey:2014yma}, which require the mediator to provide a production channel for DM at targets and are thus unlikely to be able to detect the loop portal scenario without introducing extra ingredients. For indirect detection constraints, light DMs in loop portal may annihilate into neutrinos resulting in a  relatively large number of effective relativistic degrees of freedom $N_{eff}$, compared  with $N_{eff}^{\rm SM}=3.046$~\cite{Escudero:2020dfa}, which sets a lower bound on the DM mass, $m_{\rm DM} >3.7~{\rm MeV}$ at $2\sigma$~\cite{Sabti:2019mhn}. This constraint is consistent with our parameter settings. For further discussions of impact of light DM on the Big Bang Nucleosynthesis and CMB, we refer the reader to Refs.~\cite{Sabti:2019mhn,Batell:2017rol,Iocco:2008va,Pospelov:2010hj} and references cited therein.

\section{Neutrino mass}

Integrating out the heavy Majorana neutrino $N$ and the complex scalar $\Phi$, one may get the effective Hamiltonian describing DM-neutrino interactions,
\begin{eqnarray}
{\cal H}_{\rm eff}^{A} &\approx& {{\cal I} \over m_\Phi^2} \overline{\nu} P_L^{} \chi \overline{\chi} P_R^{} \nu =  {{\cal I} \over 2 m_\Phi^2} \overline{\nu} \gamma^\mu P_L^{} \nu \overline{\chi} \gamma_\mu^{} P_R^{} \chi  \label{fierztrans}  \\ 
{\cal H}_{\rm eff}^{B} &\approx &{{\cal F}\over m_N^{} } \overline{\nu_L^{}} \nu_L^C \varphi^2 + {\rm h.c.} \label{neuanti}
\end{eqnarray}
where Fierz transformation~\cite{fierztrans} is applied to the  first equation. The Hamiltonian in Eq.~(\ref{fierztrans}) contributes to neutrino oscillation in the dark halo  which will be discussed in the next section. The Hamiltonian in Eq.~(\ref{neuanti}) contributes to neutrino Majorana mass as well as neutrino-antineutrino oscillations. In this section, we discuss the neutrino Majorana masses induced by DM in scenario B. 

Assuming  the function $f(E,~T)$ is the statistical distribution of the DM with energy $E$ and temperature $T$, one has $\int d^3 p f(E,~T) =n_\phi V$, where $n_\phi$  is the DM density of the medium, $n_\Phi V$ is the total number of the DM. 
The average of the effective Hamiltonian over the DM background in the rest frame of the medium is
\begin{eqnarray}
\int d^3 p f(E,T) \langle \varphi |\varphi^2 |\varphi\rangle {{\cal F}\over M_R^{} } \overline{\nu_L^{}} \nu_L^C =
{\rho_\varphi -3P_\varphi \over m_\varphi^2 }  {{\cal F}\over m_N^{} } \overline{\nu_L^{}} \nu_L^C \label{fakemass}
\end{eqnarray}
 where $\rho_\varphi =n_\varphi m_\varphi$ with $n_\varphi$ the DM number density   and $P_\varphi \sim n_\varphi T$ being the pressure associated with the DM.   Since $T\ll m_\varphi$, $P_\varphi$ can be safely neglected.  There are two main approaches to measure the local DM number density~\cite{Liddle:1993fq,Catena:2009mf}: local measures that use the vertical kinematics of stars near the sun; global measures which extrapolate DM density from the rotation curve.  The DM density near the sun is about $0.3\sim0.4~{\rm GeV/cm^3}$~\cite{Read:2014qva}. 
It has been shown in Refs.~\cite{Davoudiasl:2018hjw,DAmico:2018hgc} that tiny but non-zero neutrino masses may come from a long range potential sourced by DM distributions. In our  scenario B,  Majorana neutrino mass $\delta m$  can be generated in the presence of the constant DM background $\varphi$, 
\begin{eqnarray}
\delta m_\nu^{} \sim 7.68 \times 10^{-21}~({\rm eV}) \left( {{\cal F}\times \rho_\varphi^{} \over 1{\rm GeV\cdot cm^{-3}}} \right)\left( { 1~{\rm  GeV} \over m_N^{} } \right) \left(  1 ~{\rm keV} \over m_\varphi\right)^2 \; .
\end{eqnarray} 

\begin{figure}
\centering
\includegraphics[width=0.42\textwidth]{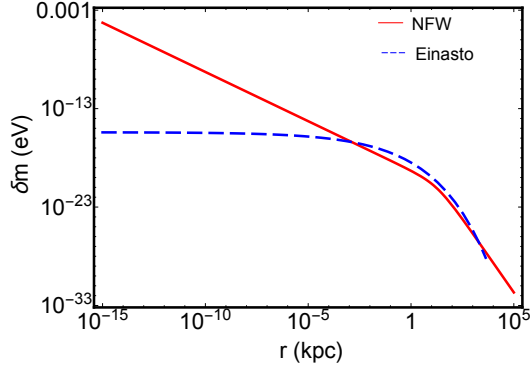}
\caption{ Neutrino screening mass as the function of the Galactic radius for NFW and Einasto DM profiles.
} \label{fig:screening}
\end{figure}

To quantify $\delta m$, we consider the DM distribution throughout the MilkyWay Galaxy.  Two popular DM density profiles are used in numerical simulations, the Navarro-Frenk-White (NFW)~\cite{Navarro:1995iw} and Einasto~\cite{Springel:2008cc}.  The distribution of the neutrino mass as a function of the Galactic radius is depicted in Fig.~\ref{fig:screening}. We take $r_\odot=8.5$ kpc, $\rho(r_\odot) =0.4 ~{\rm GeV/cm^3}$ and $R=20$ kpc for NFW profile, and take $\alpha =0.17$ and $R=20$ kpc for Einasto profile.  The red solid line represent the NFW profile and the blue dashed line represent the Einasto profile. Note that $\delta m$ is too small to give rise to any significant signals on the Earth.

%%%%%%%%%%%%%%%%%%%%%%%%%%%%%%%%%%%%%%%%%%%%%%%%%%%%%%%%%%%%%%%%%%%%%%%%%%%%%%%%%%%%%%%%%%%%%%%
%%%%%%%%%%%%%%%%%%%%%%%%%%%%%%%%%%%%%%%%%%%%%%%%%%%%%%%%%%%%%%%%%%%%%%%%%%%%%%%%%%%%%%%%%%%%%%%

\section{Neutrino oscillation}

For scenario A, the effective neutrino potential Hamiltonian in the dark halo with low temperature is ${\cal H} _{\rm eff}=V_\chi \overline{\nu_L^{} } \gamma^0 \nu_L^{} $ with the potential,
\begin{eqnarray}
V_\chi =  {{\cal I} n_\chi^{}  \over 4 m_\Phi^2}
\end{eqnarray}
where $n_\chi$ is the number density of $\chi$.  $V_\chi$ is similar to  the potential induced by the neutral current interaction of neutron. If DM interactions are neutrino flavor blinded,   $V_\chi$ can be factorized out the effective Hamiltonian describing neutrino oscillations, just like the case of neutral current interaction. Alternatively, if DM only couple to a specific neutrino flavor,  there will be matter effect induced by DM in neutrino oscillations.   

We assume DM only couples to the muon neutrino, and study the two-flavor neutrino oscillation $P(\nu_e \to \nu_\mu)$. A straight forward calculation gives   
\begin{eqnarray}
P(\nu_e \to \nu_\mu) =\left( \Delta m^2 \over \Delta \tilde m^2 \right)^2 \sin^2 2 \theta \sin^2 \left(  {\Delta \tilde m^2 x \over 4 E}\right)
\end{eqnarray}
with
\begin{eqnarray}
\Delta \tilde m^2 =\left[ (\Delta m^2 )^2 + 2\Delta m^2 \cos2\theta (A_{\rm CC}^{} - A_{\chi}^{} )+(A_{\rm CC}^{} - A_{\chi}^{} )^2\right]^{1/2}
\end{eqnarray}
where $A_{CC} =2\sqrt{2}E G_F n_e $ with $n_e$ the number density of electron in the medium and $A_\chi = {\cal I } E n_\chi / m_\Phi^2$, $\theta$ and $\Delta m^2$ are the mixing angle and the squared mass difference in the vacuum, respectively. 

For scenario B, the DM  in the medium may affect the neutrino-antineutrino oscillation probability.  We follow the quantum field theory strategy~\cite{Delepine:2009am} to  calculate the DM effect in neutrino-antineutrino oscillations.
The coefficient  in Eq.(\ref{fakemass}) plays a rule of  classical field and can modify the propagator of neutrinos,
 \begin{eqnarray}
 \left[ i\slashed{\partial} -m_\nu -\Delta(x) \right] S_A(x-x_i) = i \delta^4(x-x_i)
 \end{eqnarray}
where we have defined the coefficient in Eq.(\ref{fakemass}) as $\Delta(x)$ for simplicity.  We are unable to find a compact expression for $S_A$, but we can treat $\Delta(x)$ as a perturbation.  In this case $S_A$ can be written as
\begin{eqnarray}
S_{A}(x_f, x_i)=S_F(x_f, x_i) + \int d^4 x_1 S_F(x_f, x_1) \Delta(x_1) S_F(x_1, x_i) + \cdots 
\end{eqnarray}
Taking $\Delta $ as a constant, the propagator in the momentum space can be written as $iS_A^{-1}(p) =\slashed{p}-m_i-\Delta$, after the summation.

%%%%%%%%%%%%%%%%%%%%%%%%%%%%%%%%%%%%%%%%%%%%%%%%%%%%%%%%%%%%%%%%%%%%%%%%%%%%%%%%%%%%%%%%%%%%%%%
\begin{figure}
\centering
\includegraphics[width=0.42\textwidth]{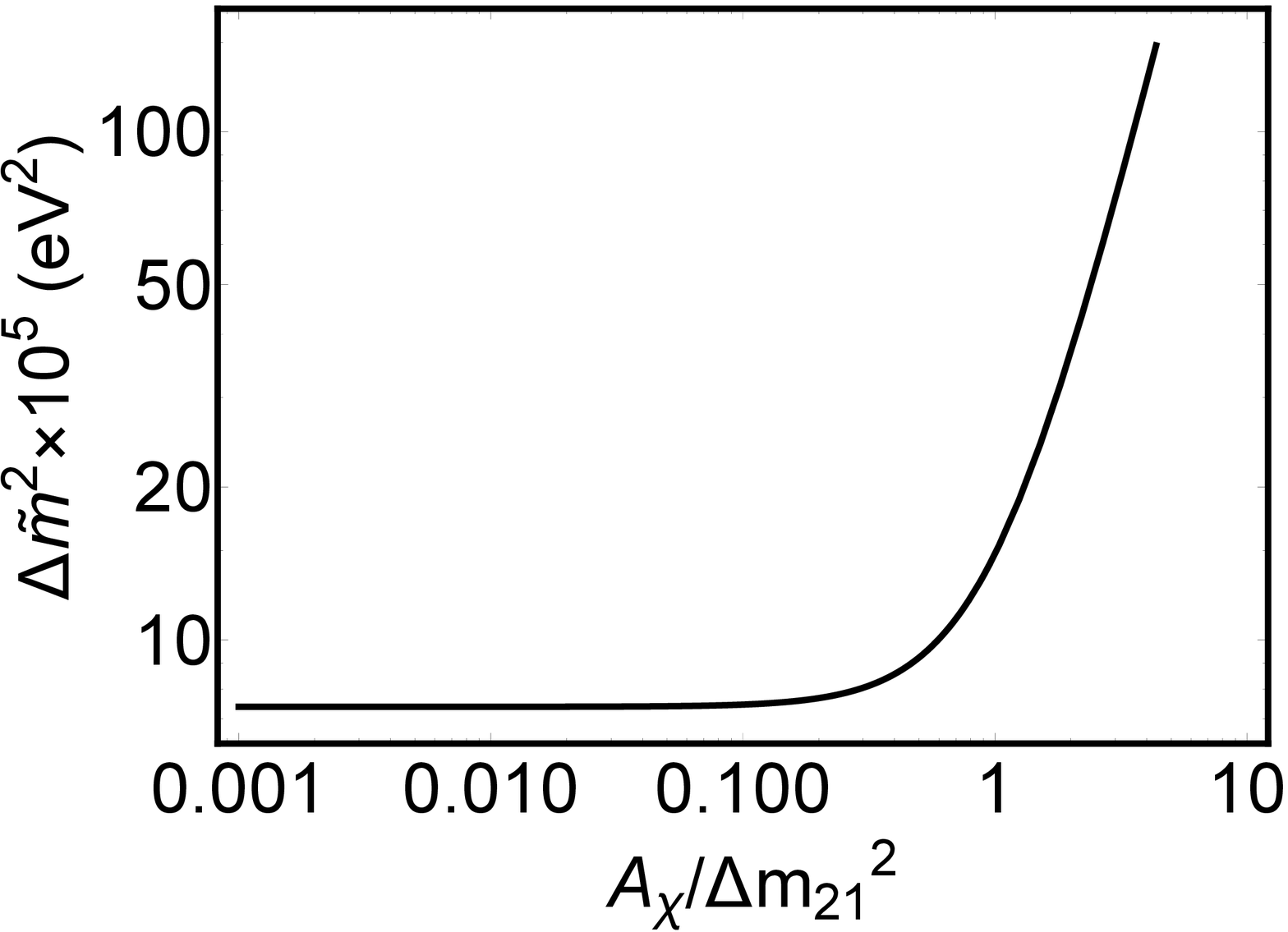}
\hspace{0.5cm}
\includegraphics[width=0.42\textwidth]{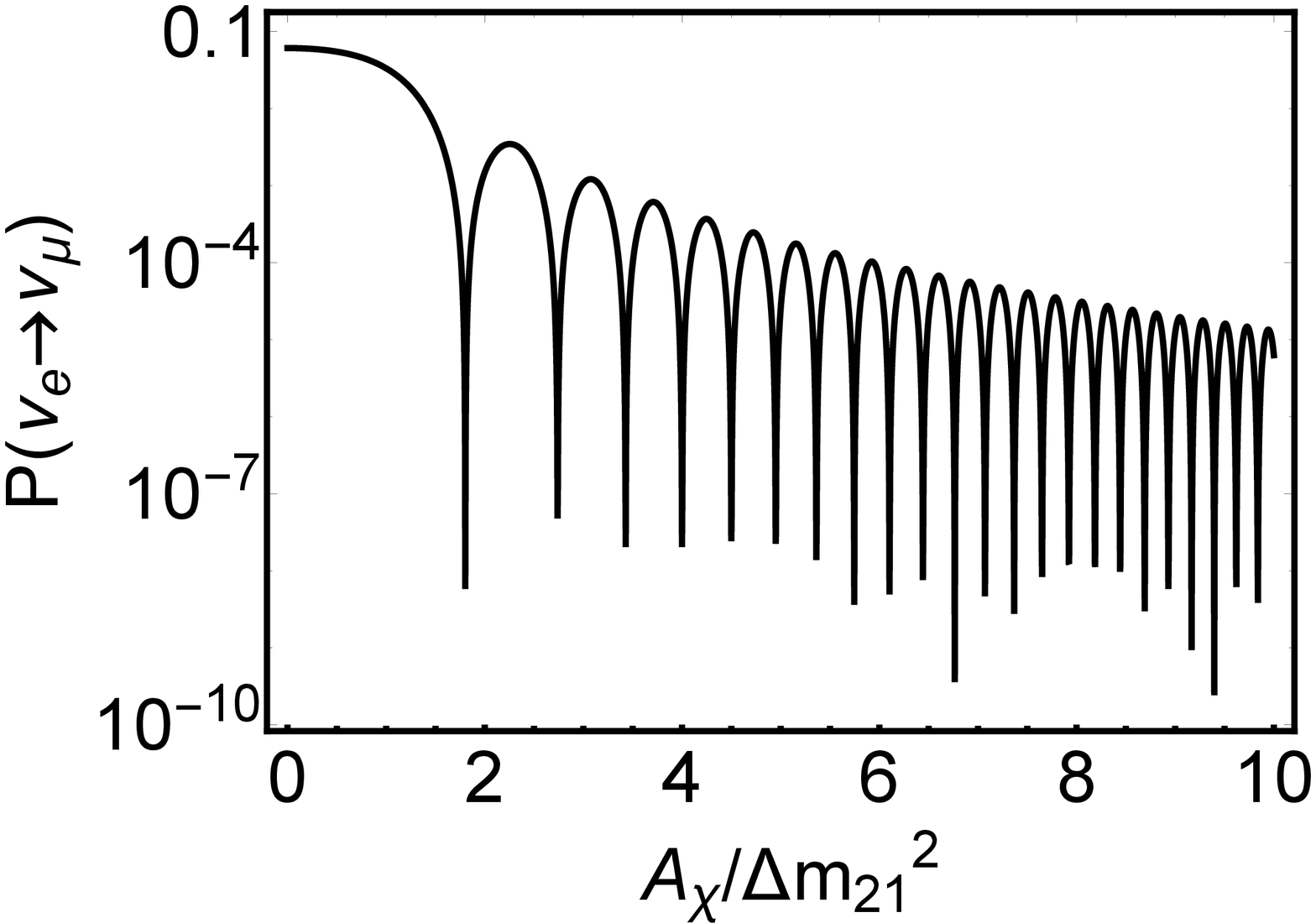}
\caption{ Left-panel: the squared mass difference $\Delta \tilde m^2$ as the function of $A_\chi/\Delta m_{21}^2$; Right-panel: the neutrino oscillation probability $P(\nu_e \to \nu_\mu)$ as the function of  $A_\chi/\Delta m_{21}^2$.
} \label{fig:oscillation}
\end{figure}
%%%%%%%%%%%%%%%%%%%%%%%%%%%%%%%%%%%%%%%%%%%%%%%%%%%%%%%%%%%%%%%%%%%%%%%%%%%%%%%%%%%%%%%%%%%%%%%

Now we can calculate the neutrino-antineutrino oscillation in the DM medium using quantum field theory approach.  The  $S$ matrix  for the process $W^-\ell_\alpha^+\to W^+ \ell_\beta^-$ is 
\begin{eqnarray}
S_{W^-\ell_\alpha^+\to W^+ \ell_\beta^-} =- \sum_i {g^2 \over 2 } U_{\alpha i }^{}  U_{\beta i}^{} (m_i + \Delta_i ) {e^{-iE_i t } \over 2 E_i } \bar u(p_f ) \slashed{\varepsilon}(k_f) \slashed{\varepsilon}(k_i) P_R^{} u(p_i) e^{ i(p_f + k_f ) t}
\end{eqnarray}
where $g$ is the weak coupling, $U$ is the so-called PMNS matrix.  As a result, neutrino-antineutrino oscillation probability  is 
\begin{eqnarray}
P(\nu_\alpha^{} \to \bar \nu_\beta^{} ) = S^2 \left| \sum_i U_{\alpha i} U_{\beta i} (m_i +\Delta_i   ){e^{-iE_i t}  \over 2 E_i }\right|^2 
\end{eqnarray}  
where $S$ is kinematic factor and is flavor blinded. This probability is same as the result of the $\nu$SM~\cite{deGouvea:2002gf,Schechter:1980gk,Li:1981um,Langacker:1998pv,Xing:2013ty} up to the replacement $m_i \to m_i +\Delta_i$.  

Note that the DM density is much smaller than that of the electron on the Earth, so the DM effect is negligible in neutrino oscillation experiments.  Here we only estimate the oscillation probability of high energy electron neutrino in a dense DM environment, where the electron number density is negligible.  $\Delta m_{21}^2$ and $P(\nu_e\to \nu_\mu)$ as the function of $A_{\chi} /\Delta m_{21}^2$ are illustrated in the left-panel and right-panel of the Fig.~\ref{fig:oscillation}, respectively. We have assumed that $x/4E=10^4 ~{\rm eV^2}$ when making the plot.  For neutrino-antineutrino oscillation, the DM effect is negligible as the Debye mass is too tiny ( see Fig.~\ref{fig:screening} for detail). However our formulae can applied to study neutrino-antineutrino oscillation in a model where there is long range force between DM and neutrino, that results in a huge Debye mass~\cite{Davoudiasl:2018hjw} in the dense DM environment. 

\section{Discussions}

Neutrino and DM are two of neutral and long-lived elementary particles that constitute the Universe.  Neutrino portal has attracted more and more attention in the past decades. In this paper, we have proposed a new scheme of DM freeze-out from the thermal bath: DM annihilates into neutrinos and freezes-out through a loop.  We showed that the annihilation cross section can be large enough to give rise to a correct relic abundance for the case where  the heavy mediator is a Majorana fermion. We further studied  the neutrino Debye mass and the matter effect in neutrino oscillations  induced by the DM. Numerical results show that the impaction of DM to neutrino properties  are negligible in our solar system. But our analytical results are applicable to investigate neutrino masses and neutrino oscillations in a dense DM environment.

\begin{acknowledgments}
The author thanks Yanyan Hu, Siyu Jiang, Mingjie Jin and Yandong Liu for helpful discussions.This work was supported by the National Natural Science Foundation of China under grant No. 11775025.
\end{acknowledgments}


\begin{thebibliography}{99}
%\cite{Aghanim:2018eyx}
\bibitem{Aghanim:2018eyx} 
  N.~Aghanim {\it et al.} [Planck Collaboration],
  %``Planck 2018 results. VI. Cosmological parameters,''
  arXiv:1807.06209 [astro-ph.CO].
  %%CITATION = ARXIV:1807.06209;%%
  %690 citations counted in INSPIRE as of 06 Apr 2019
  
  %\cite{Goldberg:1983nd,Ellis:1983ew,Servant:2002aq,Cheng:2002ej}
\bibitem{Goldberg:1983nd} 
  H.~Goldberg,
  %``Constraint on the Photino Mass from Cosmology,''
  Phys.\ Rev.\ Lett.\  {\bf 50}, 1419 (1983)
  Erratum: [Phys.\ Rev.\ Lett.\  {\bf 103}, 099905 (2009)].
  doi:10.1103/PhysRevLett.103.099905, 10.1103/PhysRevLett.50.1419
  %%CITATION = doi:10.1103/PhysRevLett.103.099905, 10.1103/PhysRevLett.50.1419;%%
  %1328 citations counted in INSPIRE as of 08 Apr 2019
  
 %\cite{Ellis:1983ew,Servant:2002aq,Cheng:2002ej}
\bibitem{Ellis:1983ew} 
  J.~R.~Ellis, J.~S.~Hagelin, D.~V.~Nanopoulos, K.~A.~Olive and M.~Srednicki,
  %``Supersymmetric Relics from the Big Bang,''
  Nucl.\ Phys.\ B {\bf 238}, 453 (1984).
  doi:10.1016/0550-3213(84)90461-9
  %%CITATION = doi:10.1016/0550-3213(84)90461-9;%%
  %1751 citations counted in INSPIRE as of 08 Apr 2019 
  
  %\cite{Jungman:1995df}
\bibitem{Jungman:1995df} 
  G.~Jungman, M.~Kamionkowski and K.~Griest,
  %``Supersymmetric dark matter,''
  Phys.\ Rept.\  {\bf 267}, 195 (1996)
  doi:10.1016/0370-1573(95)00058-5
  [hep-ph/9506380].
  %%CITATION = doi:10.1016/0370-1573(95)00058-5;%%
  %3648 citations counted in INSPIRE as of 08 Apr 2019


  
  %\cite{Servant:2002aq,Cheng:2002ej}
\bibitem{Servant:2002aq} 
  G.~Servant and T.~M.~P.~Tait,
  %``Is the lightest Kaluza-Klein particle a viable dark matter candidate?,''
  Nucl.\ Phys.\ B {\bf 650}, 391 (2003)
  doi:10.1016/S0550-3213(02)01012-X
  [hep-ph/0206071].
  %%CITATION = doi:10.1016/S0550-3213(02)01012-X;%%
  %878 citations counted in INSPIRE as of 08 Apr 2019
  
  %\cite{Cheng:2002ej}
\bibitem{Cheng:2002ej} 
  H.~C.~Cheng, J.~L.~Feng and K.~T.~Matchev,
  %``Kaluza-Klein dark matter,''
  Phys.\ Rev.\ Lett.\  {\bf 89}, 211301 (2002)
  doi:10.1103/PhysRevLett.89.211301
  [hep-ph/0207125].
  %%CITATION = doi:10.1103/PhysRevLett.89.211301;%%
  %603 citations counted in INSPIRE as of 08 Apr 2019

%\cite{Bertone:2004pz}
\bibitem{Bertone:2004pz} 
  G.~Bertone, D.~Hooper and J.~Silk,
  %``Particle dark matter: Evidence, candidates and constraints,''
  Phys.\ Rept.\  {\bf 405}, 279 (2005)
  doi:10.1016/j.physrep.2004.08.031
  [hep-ph/0404175].
  %%CITATION = doi:10.1016/j.physrep.2004.08.031;%%
  %3288 citations counted in INSPIRE as of 08 Apr 2019

  


%\cite{Essig:2011nj}
\bibitem{Essig:2011nj} 
  R.~Essig, J.~Mardon and T.~Volansky,
  %``Direct Detection of Sub-GeV Dark Matter,''
  Phys.\ Rev.\ D {\bf 85}, 076007 (2012)
  doi:10.1103/PhysRevD.85.076007
  [arXiv:1108.5383 [hep-ph]].
  %%CITATION = doi:10.1103/PhysRevD.85.076007;%%
  %313 citations counted in INSPIRE as of 22 Sep 2020

%\cite{Marsh:2015xka}
\bibitem{Marsh:2015xka} 
  D.~J.~E.~Marsh,
  %``Axion Cosmology,''
  Phys.\ Rept.\  {\bf 643}, 1 (2016)
  doi:10.1016/j.physrep.2016.06.005
  [arXiv:1510.07633 [astro-ph.CO]].
  %%CITATION = doi:10.1016/j.physrep.2016.06.005;%%
  %710 citations counted in INSPIRE as of 22 Sep 2020
  
  %\cite{Preskill:1982cy,Abbott:1982af,Dine:1981rt}
\bibitem{Preskill:1982cy} 
  J.~Preskill, M.~B.~Wise and F.~Wilczek,
  %``Cosmology of the Invisible Axion,''
  Phys.\ Lett.\  {\bf 120B}, 127 (1983).
  doi:10.1016/0370-2693(83)90637-8
  %%CITATION = doi:10.1016/0370-2693(83)90637-8;%%
  %2016 citations counted in INSPIRE as of 22 Sep 2020

%\cite{Abbott:1982af,Dine:1981rt}
\bibitem{Abbott:1982af} 
  L.~F.~Abbott and P.~Sikivie,
  %``A Cosmological Bound on the Invisible Axion,''
  Phys.\ Lett.\  {\bf 120B}, 133 (1983).
  doi:10.1016/0370-2693(83)90638-X
  %%CITATION = doi:10.1016/0370-2693(83)90638-X;%%
  %1892 citations counted in INSPIRE as of 22 Sep 2020

%\cite{Dine:1981rt}
\bibitem{Dine:1981rt} 
  M.~Dine, W.~Fischler and M.~Srednicki,
  %``A Simple Solution to the Strong CP Problem with a Harmless Axion,''
  Phys.\ Lett.\  {\bf 104B}, 199 (1981).
  doi:10.1016/0370-2693(81)90590-6
  %%CITATION = doi:10.1016/0370-2693(81)90590-6;%%
  %2643 citations counted in INSPIRE as of 22 Sep 2020



%\cite{Carr:2016drx}
\bibitem{Carr:2016drx} 
  B.~Carr, F.~Kuhnel and M.~Sandstad,
  %``Primordial Black Holes as Dark Matter,''
  Phys.\ Rev.\ D {\bf 94}, no. 8, 083504 (2016)
  doi:10.1103/PhysRevD.94.083504
  [arXiv:1607.06077 [astro-ph.CO]].
  %%CITATION = doi:10.1103/PhysRevD.94.083504;%%
  %540 citations counted in INSPIRE as of 22 Sep 2020

%\cite{Hall:2009bx}
\bibitem{Hall:2009bx}
L.~J.~Hall, K.~Jedamzik, J.~March-Russell and S.~M.~West,
%``Freeze-In Production of FIMP Dark Matter,''
JHEP \textbf{03} (2010), 080
doi:10.1007/JHEP03(2010)080
[arXiv:0911.1120 [hep-ph]].
%569 citations counted in INSPIRE as of 22 Jul 2020

%\cite{Hambye:2020lvy}
\bibitem{Hambye:2020lvy} 
  T.~Hambye, M.~Lucca and L.~Vanderheyden,
  %``Dark matter as a heavy thermal hot relic,''
  Phys.\ Lett.\ B {\bf 807}, 135553 (2020)
  doi:10.1016/j.physletb.2020.135553
  [arXiv:2003.04936 [hep-ph]].
  %%CITATION = doi:10.1016/j.physletb.2020.135553;%%
  %1 citations counted in INSPIRE as of 22 Sep 2020
  
  
  %\cite{Frigerio:2011in}
\bibitem{Frigerio:2011in}
M.~Frigerio, T.~Hambye and E.~Masso,
%``Sub-GeV dark matter as pseudo-Goldstone from the seesaw scale,''
Phys. Rev. X \textbf{1}, 021026 (2011)
doi:10.1103/PhysRevX.1.021026
[arXiv:1107.4564 [hep-ph]].
%56 citations counted in INSPIRE as of 24 Sep 2020
  

%\cite{Beneke:2014gla}
\bibitem{Beneke:2014gla}
M.~Beneke, F.~Dighera and A.~Hryczuk,
%``Relic density computations at NLO: infrared finiteness and thermal correction,''
JHEP \textbf{10} (2014), 045
doi:10.1007/JHEP10(2014)045
[arXiv:1409.3049 [hep-ph]].
%15 citations counted in INSPIRE as of 04 Sep 2020

%\cite{Beneke:2016ghp}
\bibitem{Beneke:2016ghp}
M.~Beneke, F.~Dighera and A.~Hryczuk,
%``Finite-temperature modification of heavy particle decay and dark matter annihilation,''
JHEP \textbf{09} (2016), 031
doi:10.1007/JHEP09(2016)031
[arXiv:1607.03910 [hep-ph]].
%4 citations counted in INSPIRE as of 04 Sep 2020

%\cite{Davoudiasl:2018hjw,DAmico:2018hgc}
\bibitem{Davoudiasl:2018hjw} 
  H.~Davoudiasl, G.~Mohlabeng and M.~Sullivan,
  %``Galactic Dark Matter Population as the Source of Neutrino Masses,''
  Phys.\ Rev.\ D {\bf 98}, no. 2, 021301 (2018)
  doi:10.1103/PhysRevD.98.021301
  [arXiv:1803.00012 [hep-ph]].
  %%CITATION = doi:10.1103/PhysRevD.98.021301;%%
  %10 citations counted in INSPIRE as of 22 Sep 2020

%\cite{DAmico:2018hgc}
\bibitem{DAmico:2018hgc} 
  G.~D'Amico, T.~Hamill and N.~Kaloper,
  %``Neutrino Masses from Outer Space,''
  Phys.\ Lett.\ B {\bf 797}, 134846 (2019)
  doi:10.1016/j.physletb.2019.134846
  [arXiv:1804.01542 [hep-ph]].
  %%CITATION = doi:10.1016/j.physletb.2019.134846;%%
  %6 citations counted in INSPIRE as of 22 Sep 2020

%\cite{Choi:2019zxy,Liao:2018byh,Capozzi:2018bps}
\bibitem{Choi:2019zxy}
K.~Y.~Choi, E.~J.~Chun and J.~Kim,
%``Neutrino Oscillations in Dark Matter,''
Phys. Dark Univ. \textbf{30} (2020), 100606
doi:10.1016/j.dark.2020.100606
[arXiv:1909.10478 [hep-ph]].
%4 citations counted in INSPIRE as of 22 Sep 2020

%\cite{Liao:2018byh}
\bibitem{Liao:2018byh}
J.~Liao, D.~Marfatia and K.~Whisnant,
%``Light scalar dark matter at neutrino oscillation experiments,''
JHEP \textbf{04} (2018), 136
doi:10.1007/JHEP04(2018)136
[arXiv:1803.01773 [hep-ph]].
%8 citations counted in INSPIRE as of 22 Sep 2020

%\cite{Capozzi:2018bps}
\bibitem{Capozzi:2018bps}
F.~Capozzi, I.~M.~Shoemaker and L.~Vecchi,
%``Neutrino Oscillations in Dark Backgrounds,''
JCAP \textbf{07} (2018), 004
doi:10.1088/1475-7516/2018/07/004
[arXiv:1804.05117 [hep-ph]].
%20 citations counted in INSPIRE as of 22 Sep 2020

%\cite{tHooft:1978jhc,Passarino:1978jh,Denner:1991kt}
\bibitem{tHooft:1978jhc}
G.~'t Hooft and M.~J.~G.~Veltman,
%``Scalar One Loop Integrals,''
Nucl. Phys. B \textbf{153} (1979), 365-401
doi:10.1016/0550-3213(79)90605-9
%1413 citations counted in INSPIRE as of 23 Sep 2020

%\cite{Passarino:1978jh,Denner:1991kt}
\bibitem{Passarino:1978jh}
G.~Passarino and M.~J.~G.~Veltman,
%``One Loop Corrections for e+ e- Annihilation Into mu+ mu- in the Weinberg Model,''
Nucl. Phys. B \textbf{160} (1979), 151-207
doi:10.1016/0550-3213(79)90234-7
%2489 citations counted in INSPIRE as of 23 Sep 2020

%\cite{Denner:1991kt}
\bibitem{Denner:1991kt}
A.~Denner,
%``Techniques for calculation of electroweak radiative corrections at the one loop level and results for W physics at LEP-200,''
Fortsch. Phys. \textbf{41} (1993), 307-420
doi:10.1002/prop.2190410402
[arXiv:0709.1075 [hep-ph]].
%941 citations counted in INSPIRE as of 23 Sep 2020

%\cite{Steigman:1984ac}
\bibitem{Steigman:1984ac}
G.~Steigman and M.~S.~Turner,
%``Cosmological Constraints on the Properties of Weakly Interacting Massive Particles,''
Nucl. Phys. B \textbf{253} (1985), 375-386
doi:10.1016/0550-3213(85)90537-1
%381 citations counted in INSPIRE as of 23 Sep 2020



%\cite{Chao:2019lhb,Li:2020pfy}
\bibitem{Chao:2019lhb}
W.~Chao,
%``Direct detections of Majorana dark matter in vector portal,''
JHEP \textbf{11} (2019), 013
doi:10.1007/JHEP11(2019)013
[arXiv:1904.09785 [hep-ph]].
%6 citations counted in INSPIRE as of 24 Sep 2020

%\cite{Li:2020pfy}
\bibitem{Li:2020pfy}
T.~Li and J.~Liao,
%``Loop effect in the detection of neutrinophilic dark matter model with a spin-0 mediator,''
[arXiv:2008.00743 [hep-ph]].
%0 citations counted in INSPIRE as of 24 Sep 2020


%\cite{deNiverville:2012ij}
\bibitem{deNiverville:2012ij}
P.~deNiverville, D.~McKeen and A.~Ritz,
%``Signatures of sub-GeV dark matter beams at neutrino experiments,''
Phys. Rev. D \textbf{86} (2012), 035022
doi:10.1103/PhysRevD.86.035022
[arXiv:1205.3499 [hep-ph]].
%106 citations counted in INSPIRE as of 24 Sep 2020

%\cite{Morrissey:2014yma}
\bibitem{Morrissey:2014yma}
D.~E.~Morrissey and A.~P.~Spray,
%``New Limits on Light Hidden Sectors from Fixed-Target Experiments,''
JHEP \textbf{06} (2014), 083
doi:10.1007/JHEP06(2014)083
[arXiv:1402.4817 [hep-ph]].
%42 citations counted in INSPIRE as of 24 Sep 2020

%\cite{Escudero:2020dfa}
\bibitem{Escudero:2020dfa}
M.~Escudero Abenza,
%``Precision early universe thermodynamics made simple: $N_{\rm eff}$ and neutrino decoupling in the Standard Model and beyond,''
JCAP \textbf{05} (2020), 048
doi:10.1088/1475-7516/2020/05/048
[arXiv:2001.04466 [hep-ph]].
%18 citations counted in INSPIRE as of 24 Sep 2020


%\cite{Sabti:2019mhn}
\bibitem{Sabti:2019mhn}
N.~Sabti, J.~Alvey, M.~Escudero, M.~Fairbairn and D.~Blas,
%``Refined Bounds on MeV-scale Thermal Dark Sectors from BBN and the CMB,''
JCAP \textbf{01} (2020), 004
doi:10.1088/1475-7516/2020/01/004
[arXiv:1910.01649 [hep-ph]].
%27 citations counted in INSPIRE as of 24 Sep 2020

%\cite{Batell:2017rol}
\bibitem{Batell:2017rol}
B.~Batell, T.~Han and B.~Shams Es Haghi,
%``Indirect Detection of Neutrino Portal Dark Matter,''
Phys. Rev. D \textbf{97} (2018) no.9, 095020
doi:10.1103/PhysRevD.97.095020
[arXiv:1704.08708 [hep-ph]].
%32 citations counted in INSPIRE as of 24 Sep 2020



%\cite{Iocco:2008va,Pospelov:2010hj}
\bibitem{Iocco:2008va}
F.~Iocco, G.~Mangano, G.~Miele, O.~Pisanti and P.~D.~Serpico,
%``Primordial Nucleosynthesis: from precision cosmology to fundamental physics,''
Phys. Rept. \textbf{472} (2009), 1-76
doi:10.1016/j.physrep.2009.02.002
[arXiv:0809.0631 [astro-ph]].
%414 citations counted in INSPIRE as of 24 Sep 2020

%\cite{Pospelov:2010hj}
\bibitem{Pospelov:2010hj}
M.~Pospelov and J.~Pradler,
%``Big Bang Nucleosynthesis as a Probe of New Physics,''
Ann. Rev. Nucl. Part. Sci. \textbf{60} (2010), 539-568
doi:10.1146/annurev.nucl.012809.104521
[arXiv:1011.1054 [hep-ph]].
%157 citations counted in INSPIRE as of 24 Sep 2020

%\cite{Liddle:1993fq,Catena:2009mf}
\bibitem{Liddle:1993fq}
A.~R.~Liddle and D.~H.~Lyth,
%``The Cold dark matter density perturbation,''
Phys. Rept. \textbf{231} (1993), 1-105
doi:10.1016/0370-1573(93)90114-S
[arXiv:astro-ph/9303019 [astro-ph]].
%759 citations counted in INSPIRE as of 24 Sep 2020

%\cite{Catena:2009mf}
\bibitem{Catena:2009mf}
R.~Catena and P.~Ullio,
%``A novel determination of the local dark matter density,''
JCAP \textbf{08} (2010), 004
doi:10.1088/1475-7516/2010/08/004
[arXiv:0907.0018 [astro-ph.CO]].
%510 citations counted in INSPIRE as of 24 Sep 2020

%\cite{Read:2014qva}
\bibitem{Read:2014qva}
J.~I.~Read,
%``The Local Dark Matter Density,''
J. Phys. G \textbf{41} (2014), 063101
doi:10.1088/0954-3899/41/6/063101
[arXiv:1404.1938 [astro-ph.GA]].
%282 citations counted in INSPIRE as of 07 Sep 2020


%\cite{Navarro:1995iw}
\bibitem{Navarro:1995iw}
J.~F.~Navarro, C.~S.~Frenk and S.~D.~M.~White,
%``The Structure of cold dark matter halos,''
Astrophys. J. \textbf{462} (1996), 563-575
doi:10.1086/177173
[arXiv:astro-ph/9508025 [astro-ph]].
%4917 citations counted in INSPIRE as of 07 Sep 2020


%\cite{Springel:2008cc}
\bibitem{Springel:2008cc}
V.~Springel, J.~Wang, M.~Vogelsberger, A.~Ludlow, A.~Jenkins, A.~Helmi, J.~F.~Navarro, C.~S.~Frenk and S.~D.~M.~White,
%``The Aquarius Project: the subhalos of galactic halos,''
Mon. Not. Roy. Astron. Soc. \textbf{391} (2008), 1685-1711
doi:10.1111/j.1365-2966.2008.14066.x
[arXiv:0809.0898 [astro-ph]].
%1228 citations counted in INSPIRE as of 07 Sep 2020





















 %\cite{Patel:2015tea,Patel:2016fam}
\bibitem{Patel:2015tea} 
  H.~H.~Patel,
  %``Package-X: A Mathematica package for the analytic calculation of one-loop integrals,''
  Comput.\ Phys.\ Commun.\  {\bf 197}, 276 (2015)
  doi:10.1016/j.cpc.2015.08.017
  [arXiv:1503.01469 [hep-ph]].
  %%CITATION = doi:10.1016/j.cpc.2015.08.017;%%
  %114 citations counted in INSPIRE as of 09 Apr 2019
  
  %\cite{Patel:2016fam}
\bibitem{Patel:2016fam} 
  H.~H.~Patel,
  %``Package-X 2.0: A Mathematica package for the analytic calculation of one-loop integrals,''
  Comput.\ Phys.\ Commun.\  {\bf 218}, 66 (2017)
  doi:10.1016/j.cpc.2017.04.015
  [arXiv:1612.00009 [hep-ph]].
  %%CITATION = doi:10.1016/j.cpc.2017.04.015;%%
  %28 citations counted in INSPIRE as of 09 Apr 2019



%\cite{Tremaine:1979we}
\bibitem{Tremaine:1979we}
S.~Tremaine and J.~E.~Gunn,
%``Dynamical Role of Light Neutral Leptons in Cosmology,''
Phys. Rev. Lett. \textbf{42} (1979), 407-410
doi:10.1103/PhysRevLett.42.407
%722 citations counted in INSPIRE as of 05 Sep 2020

%\cite{Farzan:2012sa}
\bibitem{Farzan:2012sa}
Y.~Farzan and E.~Ma,
%``Dirac neutrino mass generation from dark matter,''
Phys. Rev. D \textbf{86} (2012), 033007
doi:10.1103/PhysRevD.86.033007
[arXiv:1204.4890 [hep-ph]].
%101 citations counted in INSPIRE as of 05 Sep 2020

\bibitem{fierztrans}
M. E. Peskin and D. V. Schroeder, {\it An Introduction to quantum field theory.} , 1995.


%\cite{Delepine:2009am}
\bibitem{Delepine:2009am}
D.~Delepine, V.~Gonzalez Macias, S.~Khalil and G.~Lopez Castro,
%``QFT results for neutrino oscillations and New Physics,''
Phys. Rev. D \textbf{79} (2009), 093003
doi:10.1103/PhysRevD.79.093003
[arXiv:0901.1460 [hep-ph]].
%15 citations counted in INSPIRE as of 24 Sep 2020




%\cite{deGouvea:2002gf,Schechter:1980gk,Li:1981um,Langacker:1998pv,Xing:2013ty}
\bibitem{deGouvea:2002gf}
A.~de Gouvea, B.~Kayser and R.~N.~Mohapatra,
%``Manifest CP Violation from Majorana Phases,''
Phys. Rev. D \textbf{67} (2003), 053004
doi:10.1103/PhysRevD.67.053004
[arXiv:hep-ph/0211394 [hep-ph]].
%113 citations counted in INSPIRE as of 24 Sep 2020

%\cite{Schechter:1980gk,Li:1981um,Langacker:1998pv,Xing:2013ty}
\bibitem{Schechter:1980gk}
J.~Schechter and J.~W.~F.~Valle,
%``Neutrino Oscillation Thought Experiment,''
Phys. Rev. D \textbf{23} (1981), 1666
doi:10.1103/PhysRevD.23.1666
%222 citations counted in INSPIRE as of 24 Sep 2020

%\cite{Li:1981um,Langacker:1998pv,Xing:2013ty}
\bibitem{Li:1981um}
L.~F.~Li and F.~Wilczek,
%``PHYSICAL PROCESSES INVOLVING MAJORANA NEUTRINOS,''
Phys. Rev. D \textbf{25} (1982), 143
doi:10.1103/PhysRevD.25.143
%69 citations counted in INSPIRE as of 24 Sep 2020

%\cite{Langacker:1998pv,Xing:2013ty}
\bibitem{Langacker:1998pv}
P.~Langacker and J.~Wang,
%``Neutrino anti-neutrino transitions,''
Phys. Rev. D \textbf{58} (1998), 093004
doi:10.1103/PhysRevD.58.093004
[arXiv:hep-ph/9802383 [hep-ph]].
%41 citations counted in INSPIRE as of 24 Sep 2020

%\cite{Xing:2013ty}
\bibitem{Xing:2013ty}
Z.~z.~Xing,
%``Properties of CP Violation in Neutrino-Antineutrino Oscillations,''
Phys. Rev. D \textbf{87} (2013) no.5, 053019
doi:10.1103/PhysRevD.87.053019
[arXiv:1301.7654 [hep-ph]].
%26 citations counted in INSPIRE as of 24 Sep 2020


\end{thebibliography}
\end{document}